# Miniaturized Modules for Space Based Optical Communication

J. Edmunds[a], L. Henwood-Moroney[a], N. Hammond[a], E. Prowse[a], K. Hall[a], L. Szemendera[a], N. Davoudzadeh[b], P. Holland[a], K. Simpson[a], C. Palmer[a], L. Stampoulidis[c], P. Kean[a], M. Welch[a], E. Kehayas[a]

[a]G&H (Torquay), Broomhill Way, Torquay, Devon, TQ2 7QL, UK.
[b]G&H (Boston), 7 Oak Park Drive Bedford, MA, 01730, USA.
[c]Leo Space Photonics, 27 Neapoleos Street, Agia Paraskevi, 15341, Athens, Greece.

## ABSTRACT

We present recent progress in developing miniaturized optical receiver amplifiers for space communications.

**Keywords:** Free Space Optical Communications, Laser Communications, Space Photonics, Cubesats, Laser Transmitters, Constellations.

## 1. INTRODUCTION

In the last few years G&H have supplied optical components and subsystems into a range of spaced-based technology demonstrators, pathfinder missions and pioneering commercial ventures. Most notably, G&H have recently been involved in JAXA's laser communication demonstrator LUCAS, which employs a high power amplifier for transmission and low noise receiver amplifier to boost the received signal[1]. The market and in many cases governmental agencies, are transitioning beyond pathfinder missions, and looking to use optical communications as the mainstream solution for high-speed network services. The number of satellites employing or planning to employ laser terminals are growing rapidly. At the same time the size of these systems is shrinking rapidly, with commercial pressure pushing suppliers to reduce size, mass, power consumption and lead time, whilst increasing functionality and controlling cost and quality. To meet this demand and outline capability, G&H has developed miniaturized designs of transmitter and receivers for LEO laser-comms applications. These four designs are referred to in the Figure and Table below as "SmallCat", "Perseus low power (LP)", "Perseus high power (HP)" and "ORIONAS".

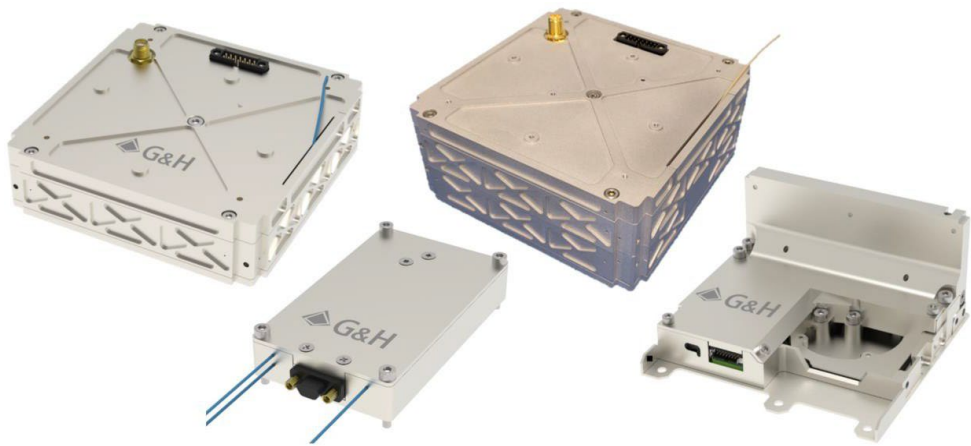

Figure 1. Rendered CAD Images and photos of G&H miniaturized optical communications modules. Top-left: 300 mW Perseus LP laser Transmitter Module. Top-Right: 3 W Perseus HP laser Transmitter Module. Bottom-Mid: ORIONAS low noise, high gain receiver amplifier Bottom-Right: SmallCat Laser Transmitter Module.

This work was supported by the H2020-SPACE-ORIONAS project from the European Union's Horizon 2020 research and innovation program under grant agreement No. 822002. This is a draft manuscript submitted to SPIE for publication. Posting of draft manuscript is permitted under SPIE sharing policies: https://www.spiedigitallibrary.org/article-sharing-policies.

The three designs of transmitters are based on amplification of a directly modulated DFB, with SmallCat and Perseus LP employing a core pumped Er amplifier stage to output > 300 mW and Perseus HP employing an additional Er/Yb amplification stage to achieve >3W output power.  The ORIONAS receiver amplifier, which is discussed in this paper, employs dual stage core pumped amplification with mid stage filtering to achieve > 50 dB of gain for a -45 dBm input. The Perseus designs are currently undergoing assembly integration and test (AIT) of engineering model (EM) builds with a view to producing a flight grade unit for a partner demonstration in 2022. A proto-flight model of the SmallCat transmitter is in AIT for an in-orbit demonstration on TNO's SmallCat platform scheduled for mid-2021 on-board NordSat-TD. An EM of the ORIONAS receiver is currently in AIT to feed into the Horizon 2020-SPACE-ORIONAS project that focuses on miniaturized space comms.

| Product Name | Description | Mass Density | Size Envelope | Power Consumption |
|---|---|---|---|---|
| SmallCat (Laser Transmitter) | Directly modulated **300 mW C-band** transmitter, for on off keying at 1Gbit/s. Direct interface to electro-optical and optoelectronic components. Bespoke form factor for TNO's SmallCat terminal. | 191 g $\approx 1.6$ g/cm$^3$ | 96x92x53 mm | < 5 W Typical for 300 mW out |
| PerseusLP (Laser Transmitter) | Directly modulated **300 mW C-band** transmitter, for on off keying at > 1G bit/s (target 10 Gbit/s). Command and Montoring through digital communication interface with onboard microcontroller, and power conditioning. Generic Cubesat form factor | 400 g $\approx 1.5$ g/cm$^3$ | 100x100x26.5 mm | 10 W Typical for 300 mW out |
| Perseus HP (Laser Transmitter) | Directly modulated **3 W C-band** transmitter, for on off keying at > 1 Gbit/s (target 10 Gbit/s). Command and Montoring through digital communication interface with onboard microcontroller, and power conditioning. Generic Cubesat form factor | 650 g $\approx 1.6$ g/cm$^3$ | 100x100x42 mm | 35 W Typical for 3 W out |
| ORIONAS (Reciever amplfier) | C-band low noise receiver amplifier providing > 50 dB of gain and ~ 5 dB Noise figure for -45 dBm input. Command and Montoring through digital communication interface with onboard microcontroller, power conditioning. Credit card sized form factor | 100 g $\approx 1.3$ g/cm | 90x54x15.5 mm | < 5 W Typical |

Table 1. Summary of the miniaturized optical communications modules designs outlined in this paper. Density is calculated with respect to space claimed by the module and, size envelope being the maximum extend of the module in each axis. All modules were designed for low earth orbit operation.

## 2. MODULE DESCRIPTION AND TESTING

Figure 2 shows a system level schematic of the ORIONAS module. Except for the fiber optic and opto-electronic components, the small form factor housing incorporates power conditioning and command and control through a microcontroller.

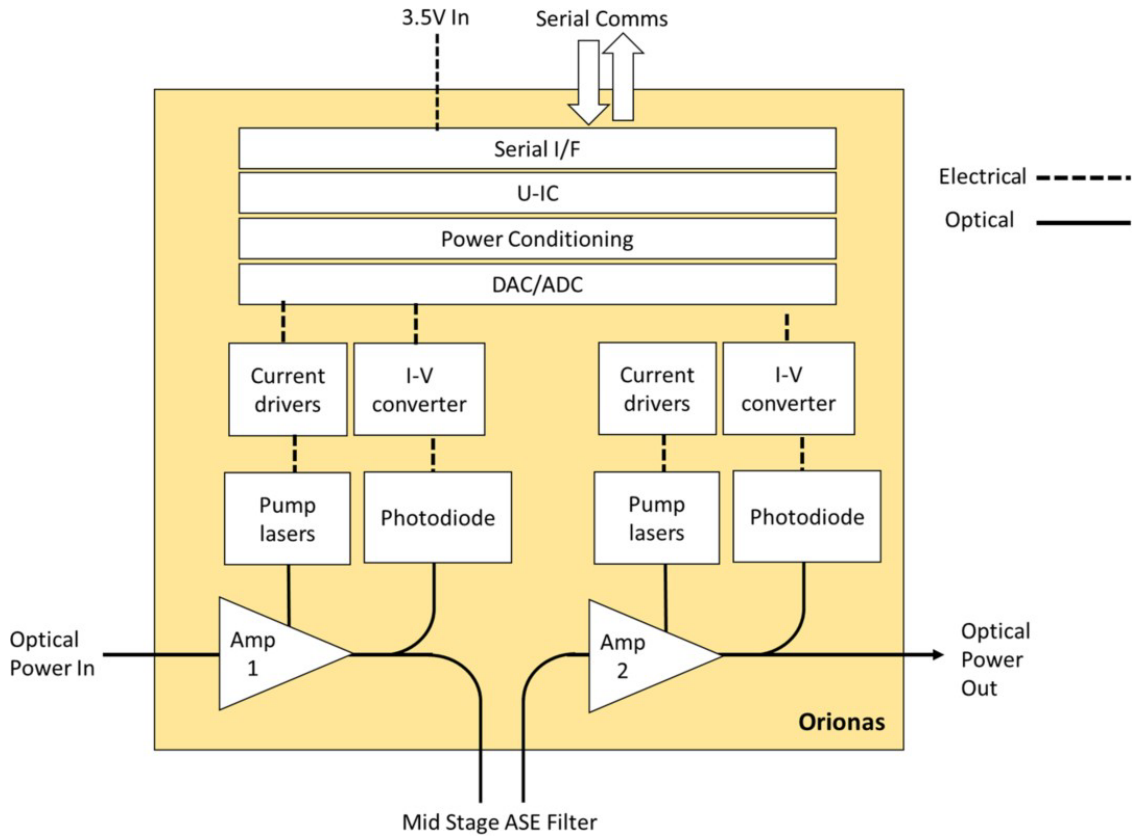

Figure 2. A simplified system level diagram of ORIONAS pre-amplifier design.

A high-efficiency and low-noise amplifier is required at the receiver end to close the inter-satellite link or uplink in case of direct downlinks with optical uplink capabilities. With a -45 dBm input signal, the ORIONAS amplifier demonstrates >50 dB gain across the C-band. To achieve this high level of gain, two separate amplifications stages are required. Spectra filtering between the amplification stages to reject out of band ASE passing is key to an efficient system; without this the second amplification stage gain is dominated by the amplification of ASE – particularly for low input power levels.

Figure 3 shows example results with 100 GHz bandwidth ASE filters, one can observe that with filtering gain is increased, particularly for low input powers. One can also observe that the filter reduces wavelength dependent gain. Due to the low-loss optical architecture, excellent noise figures of ~5 dB are delivered across the C-band. The impressive form-factor of the receiver is enabled through the use of specialized fibre and fibre-based components with a reduced cladding diameter. This allows for a smaller bend-radius to be realized for the same level of static loading, meaning the fibre can be coiled tighter without increasing the risk of fatigue failure over the unit lifetime.

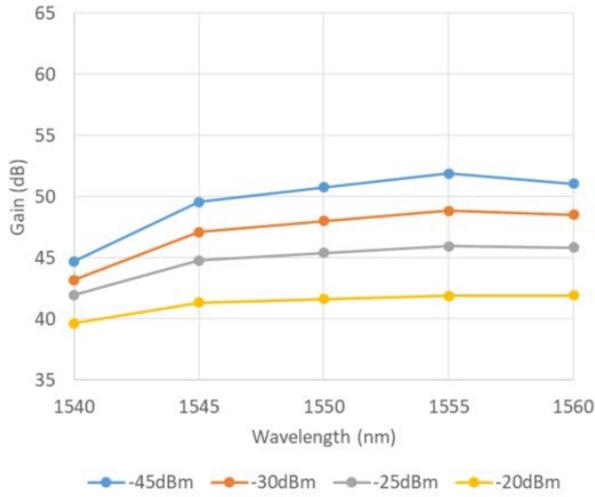
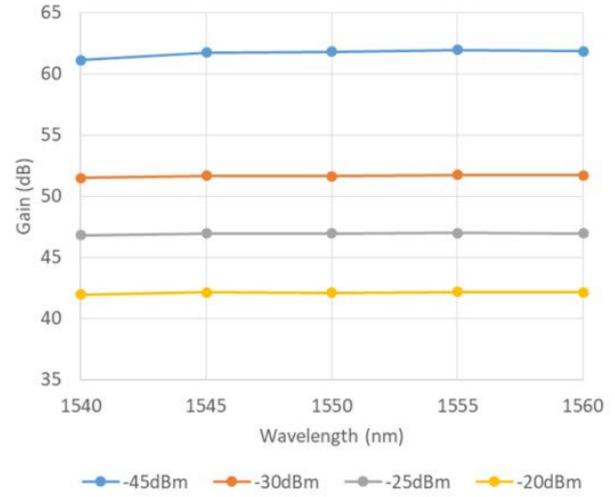
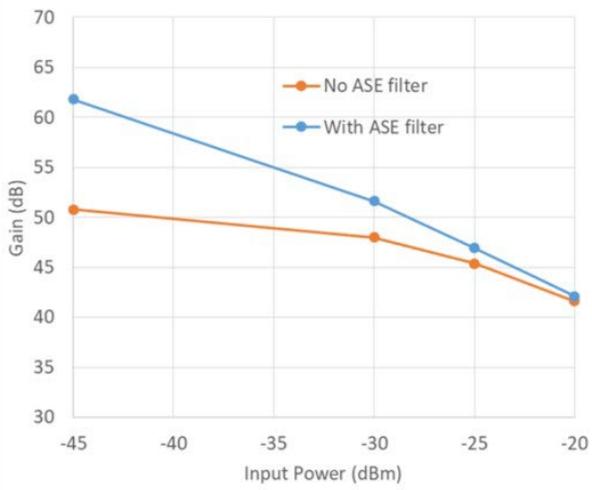
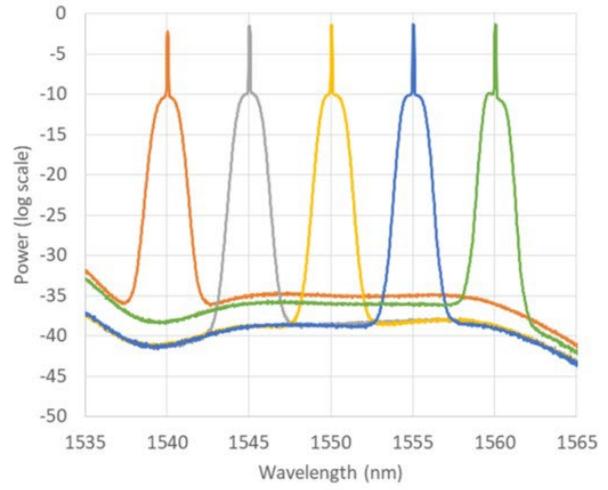

Figure 3. Top-left & Top-Right: Gain of Orionas amplifier as a function of input power and wavelength, with (right) and without (left) a 100 GHz bandwidth filter between the amplification stages, Bottom-Left: Performance as a function of input power, with and without ASE filter, for a 1550 nm signal. Bottom-Right: Example output spectra for a -45 dBm input signal at 1540, 1545, 1550, 1555 and 1560 nm with mid-stage ASE filtering employed.

# 3. ENVIROMENTAL ROBUSTNESS

The typical approach taken when qualifying a system for space (or indeed any environment), is to first select components or subsystems that have a quantified and suitable operating range. The integration of these component subsystems, then must ensure operation within these operating ranges whilst minimizing size, mass and manufacturing cost.

The commercialization of the free-space optical communications market and the emergence of New Space has set a precedent for cost-effective photonic solutions. Select COTS components are used across the outline designs, leveraging G&H's heritage in hi-reliability and space-grade photonic products[2]. With the exception of the digital and analogue electronics, all other COTS components employed have undergone environmental testing to understand their suitability for space environments[3-7].

## 3.1 Mechanical Assembly Design

Three principle environmental key challenges exist when designing a photonic module suitable for a space: radiation shielding, heat dissipation and mechanical robustness. Each come with their own impact on the size, mass, and cost of the unit.

Miniaturizing designs necessitates an increase in density, owing to the smaller volume. Most assemblies for pathfinder-type missions have densities between 0.8 - 1.1 g/cm$^3$. The designs outlined here all have densities > 1.3 g/cm3, with the SmallCat and Perseus HP having a density of ≈ 1.6 g/cm$^3$ (Table 1) – highlighting the highly compact nature of the designs. This increase in density generally aids mechanical robustness as vibrational modes are pushed to higher frequencies with the assembly becoming stiffer. The increased density also aids radiation shielding as closely located components can share shielding. There is a minimum aluminum shielding provided by the assembly wall that varies with design between 1.5 mm and 3 mm, with select localized internal shielding; this is sufficient for low earth orbit type applications.

To create optimized designs, finite element analysis (FEA) is a key design tool that can be used to quickly assess the thermal/vibrational effects of mechanical changes. This enables one to iterate the design to reduce size and mass whilst maximizing mechanical robustness and thermal performance. Section 3.2 shows example vibrational analysis results for the ORIONAS assembly.

## 3.2 Vibrational Analysis

Vibrational analysis was performed to optimise the use of mass and fixing locations. Figure 4 shows select results. The resonant frequencies simulated are much higher than the main excitational bands generated by launchers, which are typically peak at frequencies < 500 Hz. Furthermore all fundamental modes correspond to movement of surfaces onto which critial components aren't mounted.

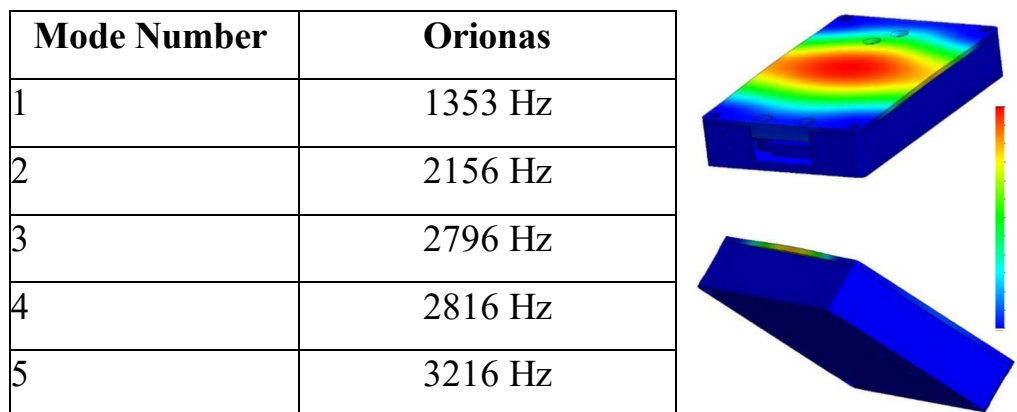

| Mode Number | Orionas |
|---|---|
| 1 | 1353 Hz |
| 2 | 2156 Hz |
| 3 | 2796 Hz |
| 4 | 2816 Hz |
| 5 | 3216 Hz |

Figure 4. (left) First five computed vibrational modes of the ORIONAS assembly. The high vibrational frequencies indicate very stiff assembly. These values where computed from the metalwork as select components only, with most housed componentry being added as non-structural mass to produce a worse case analysis. (right) Visualizations from two angles of the lowest vibrational modes of the ORIONAS assembly. The colour scale is linear and arbitrarily scaled for each assembly to best visualise the vibrational mode.

## 4. CONCLUSION

This paper highlighted our recent development work in miniaturizing optical receiver amplifiers for space communications.

## ACKNOWLEDGEMENT

This project has received funding from the European Space Agency's ARTES programme "Miniaturized Laser-Comm Transmitters for Small Satellite Platforms" and the European Union's Horizon 2020 research and innovation programme under the Grant Agreement No. 822002. We would like to thank colleagues from the Free-Space Optical Communications team at TNO for the fruitful collaboration.